# Suppressing parasitic flow in membraneless diffusion-based microfluidic gradient generators

Vahid Khandan,[1] Ryan C. Chiechi,[2,3] Elisabeth Verpoorte[1] and Klaus Mathwig*,[1,4]

[1] *University of Groningen, Groningen Research Institute of Pharmacy, Pharmaceutical Analysis, 9700 AD Groningen, the Netherlands.*

[2] *Stratingh Institute for Chemistry and Zernike Institute for Advanced Materials, University of Groningen, Nijenborgh 4, 9747 AG Groningen, the Netherlands*

[3] *Department of Chemistry & Organic and Carbon Electronics Laboratory, North Carolina State University, Raleigh, NC, 27695 USA*

[4] *imec within OnePlanet Research Center, Bronland 10, 6708 WH Wageningen, the Netherlands*

*Email: klaus.mathwig@imec.nl

5 November 2024

Diffusion-based microfluidic gradient generators (DMGGs) are essential for various in-vitro studies due to their ability to provide a convection-free concentration gradient. However, these systems, often referred to as membrane-based DMGGs, exhibit delayed gradient formation due to the incorporated flow-resistant membrane. This limitation substantially hinders their application in dynamic and time-sensitive studies. Here, we accelerate the gradient response in DMGGs by removing the membrane and implementing new geometrical configurations to compensate for the membrane's role in suppressing parasitic flows. We introduce these novel configurations into two microfluidic designs: the H-junction and the Y-junction. In the H-junction design, parasitic flow is redirected through a bypass channel following the gradient region. The Y-junction design features a shared discharge channel that allows converging discharge flow streams, preventing the buildup of parasitic pressure downstream of the gradient region. Using hydraulic circuit analysis and fluid dynamics simulations, we demonstrate the effectiveness of the H-junction and Y-junction designs in suppressing parasitic pressure flows. These computational results, supported by experimental data from particle image velocimetry, confirm the capability of our designs to generate a highly stable, accurate, and convection-free gradient with rapid formation. These advantages make the H-junction and Y-junction designs ideal experimental platforms for a wide range of in-vitro studies, including drug testing, cell chemotaxis, and stem cell differentiation.

## Introduction

Generating a concentration gradient, though seemingly straightforward, requires a well-devised strategy to ensure precise control over the established microenvironment. This is particularly critical for in-vitro studies, where concentration gradients regulate key cellular processes such as proliferation, differentiation, and chemotaxis.[1–5] To achieve this, microfluidic gradient generators (MGGs) have emerged as potent tools, offering remarkable temporal stability and spatial precision[6–9] compared to traditional gradient generators such as micropipettes and chamber systems.[10–13] These advantages, combined with cost-effective fabrication, high-throughput capabilities, and reduced sample consumption, make MGGs indispensable in biomedical studies such as drug testing and apoptotic assays.devre[10,12,14,15]

Despite the variety in their designs,[16] MGGs generally fall along a design spectrum ranging from convection-based (CMGGs) to diffusion-based MGGs (DMGGs), distinguished by their distinct mixing mechanisms.[7,8] CMGGs use the controlled flow of solutions with varying concentrations through network designs to achieve precise control over gradient formation.[17,18] For example, in tree-like gradient generators, a series of branching channels systematically divide and merge fluid streams, generating gradients through serial dilution and mixing.[18,19] CMGGs are highly versatile, capable of generating stable gradients and allowing real-time control of gradient profiles by adjusting flow rates. Additionally, they can create complex nonlinear gradients, including exponential, sigmoidal, parabolic, and periodic concentration gradients, which sets them apart from other gradient generators.[17,20] The primary drawback of CMGGs is their inherent inability to create a convection-free microenvironment within the gradient region. The resultant convective flow induces parasitic shear stresses, which can significantly alter cellular responses to the gradient and can obscure experimental results in in-vitro studies.[7,19]

DMGGs, in contrast, leverage passive diffusion to create a concentration gradient between a high concentration source and a low concentration sink.[21,22] These systems, often referred to as membrane-based MGGs,[8,23] incorporate an interconnecting medium that typically features a flow-resistant membrane such as a biological hydrogel. This membrane, which is permeable to the substance of interest, separates the gradient region from convective flows and prevents the development of parasitic shear stress within the gradient region. The main advantage of DMGGs is their ability to effectively suppress parasitic flows, providing an in-vitro environment that closely mimics extracellular conditions. However, these systems substantially suffer from delayed gradient formation and difficulties in quickly adjusting gradient properties. The convection-free condition in DMGGs is achieved at the expense of losing the speed of gradient formation.[7,8,22] This significantly hinders the application of DMGGs in dynamic and time-sensitive studies such as chemotaxis of highly motile invasive cells, where rapid gradient formation is crucial.[2,19,24]

Suppressing parasitic flow in membraneless diffusion-based microfluidic gradient generators

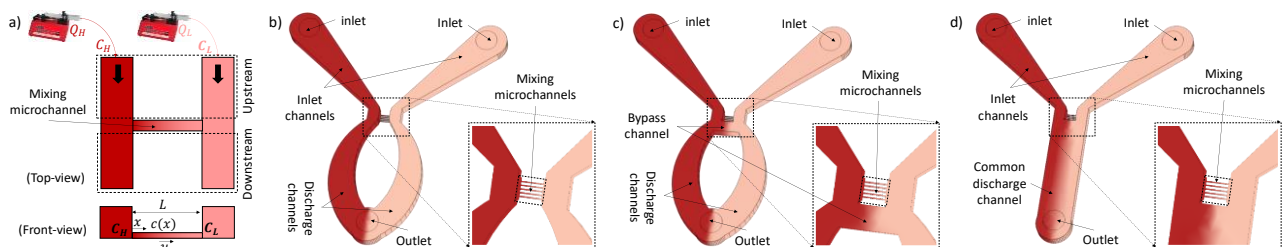

**Fig. 1 a)** Schematic representation of a DMGGs, where passive diffusion within the mixing microchannel creates a concentration gradient $c(x)$, dictated by the boundary conditions $C_H$ and $C_L$ and parasitic flow $u$. **b)** Reference design without mechanisms to mitigate parasitic pressure, consisting of two inlet channels leading to an array of mixing microchannels and branching out into two discharge channels. **c)** New H-junction design with a bypass channel immediately following the mixing microchannels, aimed at redirecting parasitic flow. **d)** New Y-junction design further advanced by incorporating a shared discharge channel, which allows for the dissipation of parasitic pressure as two discharge streams converge.

One practical way to address this technical trade-off in DMGGs is to replace the traditional flow-resistant membrane with a mixing microchannel connecting the sink and source side channels (see Fig. 1a).[19] Although such a membraneless DMGG facilitates rapid gradient formation, it remains prone to the development of parasitic flows due to the lower hydraulic resistance of the mixing microchannel compared to the membrane.[25]

Here, we address this issue by introducing two new designs in the present study: the H-junction and the Y-junction (see Fig. 1c, d). Our designs effectively manage pressure differences between the sink and source side channels, thereby minimizing parasitic flows within the mixing microchannel. The H-junction design incorporates a bypass channel, which effectively redirects parasitic flows away from the mixing microchannel and releases excess pressure. The Y-junction, on the other hand, features a shared discharge channel, allowing for smooth flow convergence to reduce pressure differentials. Both configurations are positioned downstream of the mixing microchannel, ensuring that parasitic pressure is dissipated and the associated parasitic flow and shear stress within the gradient region are suppressed.

To validate the effectiveness of these designs, we conducted comprehensive simulations and experimental assessments. Using hydraulic circuit analysis and numerical finite element simulations, we successfully modelled the suppression of parasitic pressure flows and the increased precision in the established gradients achieved by the H-junction and Y-junction designs. Additionally, we fabricated prototypes of our membraneless DMGGs and employed Particle Image Velocimetry (PIV) to visualize the suppression of parasitic flows and validate the simulation results.

These combined computational and experimental efforts demonstrate that our innovative designs not only accelerate the gradient formation process in DMGGs, but also ensure the maintenance of a convection-free concentration gradient.

Importantly, these designs offer significant advantages in fabrication simplicity. Unlike alternative devices,[25–32] including gradient chambers connected to sink/source side channels through very shallow microchannels,[33] our designs are more convenient to implement, reducing both fabrication time and cost. The enhanced reproducibility, stability, precision, along with an accelerated gradient response, make the H-junction and Y-junction designs ideal experimental platforms for a wide range of in-vitro studies.

## Materials and Methods

### Materials

Carboxylated polystyrene nanoparticles (FluoSpheres), with a nominal diameter of 200 nm and excitation/emission wavelengths of 505/515 nm, were sourced from Thermo Fisher Scientific (Eugene, Oregon, USA) under the catalogue number F8811 and stored at 4°C. The reagents required for surface modification, including succinic anhydride (SA), (3-aminopropyl)triethoxysilane (APTES), absolute ethanol, and dimethylformamide (DMF), were acquired from Sigma-Aldrich (NL). Poly(dimethylsiloxane) (PDMS) from Dow Corning (Sylgard 184) and SU8 negative photoresists (SU8-2002 and SU8-2100) from MicroChem (Newton, MA) were employed for soft lithography and photolithography, respectively. A Harrick plasma cleaner (USA) was used for surface activation. Fluid flow within the devices was driven by syringe pumps from Harvard Apparatus, with glass micro-syringes provided by ILS GmbH (Germany), and Tygon tubing featuring inner and outer diameters of 0.25/0.76 mm, sourced from Avantor VWR (USA).

### Device Fabrication and Surface Treatment

The fabrication of microfluidic devices was achieved through a replica molding process.[34,35] Initially, the 3D-inverted features of the microfluidic chips were patterned onto a two-layer master mold using photolithography inside a cleanroom. The first layer featured shallow mixing microchannels with a height of 2 μm, fabricated using SU-8 2002 negative photoresist.

The second layer, designed to include the remaining structures of inlet and outlet side channels and the bypass channel, was 200 μm thick and made with SU-8 2100 negative photoresist. Both layer designs were structured on a chromium photomask from DeltaMask (the Netherlands). Subsequent steps, including pre-baking, exposure, post-baking, and hard-baking, followed MicroChem's data sheet guidelines,[36,37] with the exception of the post-baking of SU-8 2100, which was performed at 65°C for 15 minutes in a single step.

Following the fabrication of the master mold, the 3D-inverted features were transferred to PDMS using soft lithography performed outside the cleanroom. A mixture of PDMS prepolymer and curing agent in a 10:1 weight ratio was gently poured onto the mold and cured on a perfectly levelled hotplate at 70°C for two hours. After curing, the PDMS layer was peeled off, the devices were separated by cutting, and 1 mm inlet and outlet holes were created using a biopsy punch. The PDMS layer and glass coverslip were then cleaned with an air gun and scotch tape before being activated with oxygen plasma for 25 s at a pressure of 310-320 mTorr. The activated surfaces were bonded immediately without pressure and placed on a hotplate at 70°C for 2 minutes to enhance the glass-PDMS bonding.

To functionalize the interior surfaces of the microfluidic device with carboxyl groups, a stepwise surface modification process was employed.[38–41] This surface functionalization is crucial for preventing the adsorption of carboxylate nanoparticles. Initially, a 5% solution of APTES in absolute ethanol was introduced right after the PDMS-glass bonding to react with the hydroxyl groups formed during plasma activation, leaving amine groups on the surface. After 30 minutes, the silane solution was flushed out, and the chips were gently rinsed with absolute ethanol followed by DMF. Next, a 25 mM solution of succinic anhydride in DMF was introduced to react with the $NH_2$ groups, forming carboxyl groups. The chips were heated on a hotplate at 100°C for 2 hours, then flushed with DMF to remove the SA solution. Finally, the treated microfluidic chips were rinsed with milli-Q water, and the inlet and outlet connections were sealed with scotch tape to preserve the modified surfaces for subsequent optical measurements.

Before conducting optical measurements, sodium hydroxide was added to the suspension of carboxylated nanoparticles to reach a concentration of 10 mM NaOH. This step deprotonated the carboxyl groups, resulting in negatively charged surfaces on both the nanoparticles and the interior of the microchannels. The negative charge prevents aggregation and adsorption of the nanoparticles due to electrostatic repulsion, thereby improving the accuracy of the optical measurements.

### Fluorescence Microscopy

High-resolution fluorescence microscopy was conducted using a DeltaVision Elite system (GE Healthcare UK Ltd., Little Chalfont, UK) at the Microscopy and Imaging Center of the University Medical Center Groningen. The microscope was equipped with a 60x PLAPON/1.42 oil immersion objective (Olympus) and a FITC/GFP filter set (461-489 nm excitation, 501-549 nm emission). Pixels were binned 2x2 to achieve optimal resolution of 217 nm in a region-of-interest (ROI) of 512x512 pixels while utilizing full excitation power. Time-lapse images were captured with a PCO-edge sCMOS camera and recorded at a time resolution of 100 ms using the softWoRx software (on the Linux CentOS 6.3 platform).

### Particle Image Velocimetry

The trajectory of individual 200 nm nanoparticles was extracted from the recorded time-lapse images using TrackMate,[42] a built-in plugin of the Fiji (ImageJ) software. The positions of tracers were initially localized in all frames using the Laplacian of Gaussian (LOG) detector. Subsequently, the identified positions were linked from frame to frame using the simple Linear Assignment Problem (LAP) mathematical framework to determine particle trajectories. Each identified trajectory specifies the position of a particle within the observation microchannel in the form of a time series, $r_{i=1,2,3,...,N} = [x(i\delta), y(i\delta)]$, where $\delta$ is the time resolution of the recorded time-lapse images with N frames. Afterwards, the Time-Averaged Mean Square Displacements (TAMSD) in both the $x$ and $y$ directions, denoted as $\rho_x(\tau)$ and $\rho_y(\tau)$, respectively, are calculated for different lag times $\tau = n\delta$ (where $n = 1,2,3,...,N-1$). TAMSDs of $\rho_x(\tau)$ and $\rho_y(\tau)$ are given by

$$\rho_x(\tau) = (\frac{1}{N-n}) \sum_{m=1}^{N-n} (x(m\delta + \tau) - x(m\delta))^2) \quad (1)$$

$$\rho_y(\tau) = (\frac{1}{N-n}) \sum_{m=1}^{N-n} (y(m\delta + \tau) - y(m\delta))^2).$$

Finally, the components of the average velocity vector, $(v_x, v_y)$ are obtained by fitting the model functions of $\rho_x(\tau) = 4D\tau + v_x^2\tau^2$ and $\rho_y(\tau) = 4D\tau + v_y^2\tau^2$ to the calculated TAMSDs.[43–46]

### Fluid Dynamics Simulation

Two-dimensional drawings of microfluidic device layers are initially generated using CLEWIN 3.0 software. These designs are subsequently imported into SOLIDWORKS software (version 2019) as DXF format to incorporate thickness and produce three-dimensional geometries. The resulting 3D geometries are then imported into COMSOL Multiphysics 5.5 as STEP files. Afterwards, a stationary computational model is developed by employing "Single Phase Laminar Flow" and "Transport of Diluted Species" interfaces to simulate pressure-driven flow and concentration gradient. The boundary conditions of the microfluidic device's inlets and outlet are set as constant concentrations and inflows with different flow rates, and flow discharge at atmospheric pressure. The model is subsequently meshed automatically (physics-controlled mesh) with normal size, and the results are collected using a MATLAB script. Upon completion of the simulation, this MATLAB script initiates the next round of simulation by updating the geometry and boundary conditions in the applied model.

## Results and Discussion

### Theoretical models for DMGGs with parasitic flow

In this study, the mechanism for generating gradients relies on the diffusion of substances of interest through a mixing microchannel. This mixing microchannel, positioned centrally as depicted in Fig. 1a, connects two side channels. The dimensions of the microchannel are 2 µm × 5 µm × 100 µm (height, width, and length). The side channels, significantly larger with a height of 200 µm and a width of 500 µm, are constantly filled with the solutions at varying solute concentrations, as denoted by $C_H$ and $C_L$. The essential role of these side channels is to regulate the concentration gradient, guaranteeing that the ends of the mixing microchannel preserve constant preset concentrations. The established steady-state gradient along the length of the mixing microchannels, in the $x$-direction, conforms to the convection-diffusion equation for incompressible flow, assuming no sinks and sources are present. This equation relates the concentration of the substances of interest, denoted by $c(x,t)$, to the diffusion coefficient, $D$, and parasitic flow velocity, $u$, towards the gradient direction as[47]

$$\frac{\partial c}{\partial t} = \frac{\partial}{\partial x}(uc) + D\frac{\partial^2}{\partial x^2}c. \quad (2)$$

For a constant parasitic flow velocity and with boundary conditions of constant concentrations as illustrated in Fig. 1a, the steady state solution can be expressed as

$$c(x^*) = C_H - \frac{(C_H - C_L)}{(1 - e^{-\text{Pé}})}(1 - e^{-\text{Pé}x^*}). \quad (3)$$

Here, $x^* = x/L$, where $L$ denotes the length of the microchannel, and $\text{Pé} = uL/D$ is the dimensionless Péclet number. The Péclet number compares the magnitude of convective transport, described by the convection time along the channel $L/u$, with diffusive transport, represented by diffusion time $L^2/D$. A Péclet number considerably greater than unity ($\text{Pé} \gg 1$) indicates that convection is the dominant transport mechanism, whereas a negligible Péclet number ($\text{Pé} \approx 0$) signifies a dominance of diffusion.

Concentration gradients, derived from eqn. (3), can be normalized with respect to the applied boundary conditions, leading to the definition of the dimensionless gradient profile $c^*$, given by

$$c^*(x^*) = \frac{C_H - c(x^*)}{C_H - C_L} = \frac{1 - e^{-\text{Pé}x^*}}{1 - e^{-\text{Pé}}}. \quad (4)$$

As shown in Fig. 2a, in an ideal purely diffusion-driven environment ($\text{Pé} = 0$), a linear gradient profile, $c^*_{\text{id}}(x^*) = x^*$, is established. This condition is typically preferred as it facilitates the straightforward determination of local concentrations across the gradient domain, directly from the preset boundary concentrations of $C_H$ and $C_L$. However, the presence of parasitic flow disrupts this ideal linear profile, resulting in a nonzero Péclet number and, consequently, deviating from the preferred linear profile ($c^*_{\text{id}}$). Fig. 2b depicts this deviation as $\delta c^*(x^*) = c^* - c^*_{\text{id}}$, within the gradient region. This profile deviation, which intensifies with an increasing Péclet, complicates the analysis of processes involving the concentration gradient, since it requires the evaluation of Péclet to ascertain the local concentration. Therefore, minimizing parasitic flow to achieve a zero Péclet number in diffusion-based gradient generators is crucial for ensuring high accuracy, predictability, and reproducibility of experiments.

Moreover, the gradient profile deviation, $\delta c^*$, (see Fig. 2b) varies across the gradient region, exhibiting a local peak whose position and magnitude are also dependent on Pé. To quantitatively assess this spatial variation, the root-mean-square deviation of $\delta c^*$ can be used as a dimensionless parameter

$$\delta = \sqrt{\int_0^1 \delta c^{*2} dx^*} = \sqrt{\int_0^1 [c^*(x^*) - x^*]^2 dx^*}. \quad (5)$$

This parameter, $\delta$, quantifies the deviation from the ideal linear gradient and can be analytically derived from eqn. (5) by incorporating eqn. (4), resulting in

$$\delta = \sqrt{\frac{1}{(1 - e^{-\text{Pé}})^2} - \frac{\left(1 + \frac{3}{\text{Pé}}\right)}{1 - e^{-\text{Pé}}} + \left(\frac{1}{3} + \frac{3}{2\text{Pé}} + \frac{2}{\text{Pé}^2}\right)}. \quad (6)$$

As illustrated in Fig. 2c, parameter $\delta$ yields an intuitive visualization of the deviation in the gradient profile for $\text{Pé} > 0$. As a result, it acts as a characteristic parameter reflecting the gradient generator's response to the developed parasitic flow, it is an indicator for the mitigation of parasitic flows. Additionally, Fig. 2d shows that the $\delta$ values correlate with specific Pé values, introducing a novel method for quantifying Pé without relying on fluid velocimetry or diffusivity measurements. More importantly, the $\delta$ parameter can be reported independently of channel geometry to ascertain the maximum allowable Péclet number ($\text{Pé}_{\max}$) for a given experiment, based on the sensitivity of empirical data to the gradient profile. This approach allows for considering specific $\text{Pé}_{\max}$ thresholds that align with the acceptable level of deviation from the ideal gradient profile, ensuring the integrity and reliability of the experimental outcomes.

## H-junction and Y-junction Designs

In this study, we enhanced the design of DMGGs by introducing features that minimize parasitic pressure flow and the resulting shear stress within the gradient region. These improvements focus on controlling parasitic pressure, which primarily arises from geometric mismatch and flow imbalances, such as varying flow rates between downstream side channels of the mixing microchannels (as shown in Fig. 1a). By modifying the downstream region, we achieved a substantial reduction in parasitic pressure within the mixing microchannel. Building on this concept, we introduce two new designs to minimize parasitic flow within the gradient domain:

1. The *H-junction* design: This design incorporates a bypass channel with significantly lower hydraulic resistance than the mixing microchannel, positioned immediately after the gradient region (as shown in Fig. 1c).
2. The *Y-junction* design: In this design, a shared discharge channel replaces individual discharge

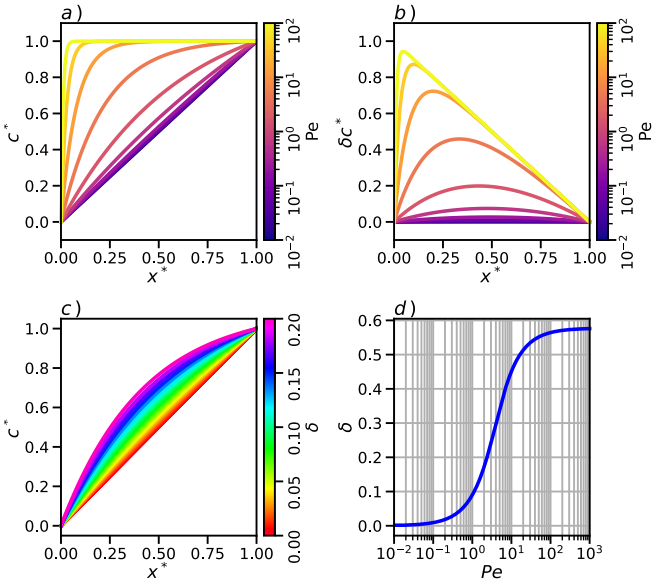

**Fig. 2.** Demonstration of concentration gradients $c(x)$ in DMGGs influenced by parasitic flows. **a)** Normalized concentration gradients between the boundary conditions $C_H$ and $C_L$, as defined by eqn. (4) **b)** Deviation in gradient profiles induced by parasitic flow (Pé > 0) compared to the ideal linear distribution in a purely diffusion-driven system (Pé = 0). **c)** Gradient profiles corresponding the parameter $\delta$, as defined in eqn. (5). **d)** Analytical depiction of the relationship between $\delta$ and the Péclet number, following the formulation presented in eqn. (6).

Suppressing parasitic flow in membraneless diffusion-based microfluidic gradient generators channels, allowing the combined flow of two discharge streams toward the outlet. This configuration is depicted in Fig. 1d.

Both approaches build on the principle of parallel flow control.[48,49] Specifically, a larger bypass or common discharge channel is placed parallel to a mixing channel with a smaller cross-section and much lower hydraulic resistance. This arrangement reduces flow in the mixing channel proportionally to the ratio of hydraulic resistances between the two parallel channels.[50]

To evaluate the effectiveness of these designs, we also use a *Reference* design that shares the same geometry but lacks specialized features to mitigate pressure differences. This baseline design, shown in Fig. 1a and b, serves as a benchmark for comparison with the H-junction and Y-junction designs.

### Hydraulic Circuit Analysis

Fig. 3 presents the equivalent hydraulic circuit diagrams for the Reference, H-junction, and Y-junction designs, illustrating the mechanisms to minimize parasitic pressure within the mixing microchannels. In these diagrams, syringe pumps act as controllable fluid-flow sources, precisely introducing solutions at constant flow rates. The system components—microfluidic channels and inlet/outlet tubing—are modelled as hydraulic resistors, each impeding the flow passing through them. All inlets and outlets are set to zero potential, reflecting the practical condition where syringe pumps and output flows operate at atmospheric pressure.

The critical metric for evaluating the effectiveness of the design modifications in reducing parasitic flow is the parasitic pressure within the mixing microchannels, measured as the pressure difference between nodes A and B (see Fig. 3). This parameter is calculated for the Reference ($\Delta P_R$), H-junction ($\Delta P_H$), and Y-junction ($\Delta P_Y$) designs from hydraulic circuit analysis as follows:

- $\Delta P_R = R_M \left( \frac{R_H Q_H - R_L Q_L}{R_M + R_H + R_L} \right)$,
- $\Delta P_H = \left( \frac{R_M R_B}{R_M + R_B} \right) \left( \frac{R_H Q_H - R_L Q_L}{\frac{R_M R_B}{R_M + R_B} + R_H + R_L} \right)$, (7)
- $\Delta P_Y = 0$.

Here, $Q_H$ and $Q_L$ (as illustrated in Fig. 3) represent the inflow rates in the high and low-concentration inlet channels, driven by syringe pumps. The parameters $R_M$, $R_B$, $R_H$, and $R_L$ denote the hydraulic resistances of the mixing channel, bypass channel, and high- and low-concentration discharge channels, respectively.

To generalize the analysis and make it independent of specific system geometry, fluid properties, and experimental conditions, we define the dimensionless parameters $Q_R = Q_H/Q_L$, referred to as the imbalanced flow, and $R_D = R_H/R_L$, termed as the discharge mismatch. Additionally, the hydraulic resistance of the mixing and bypass channels can be normalized as $R_M^* = R_M/R_L$, and $R_B^* = R_B/R_L$, respectively. Using these dimensionless parameters, the eqn. (7) for the Reference and H-junction designs are given by:

- $\Delta P_R^* = R_M^* \left( \frac{R_D Q_R - 1}{R_M^* + R_D + 1} \right)$,
- $\Delta P_H^* = \left( \frac{R_M^* R_B^*}{R_M^* + R_B^*} \right) \left( \frac{R_D Q_R - 1}{\frac{R_M^* R_B^*}{R_M^* + R_B^*} + R_D + 1} \right)$, (8)

In this dimensionless expression, $\Delta P_R^*$ and $\Delta P_H^*$ represent the normalized parasitic pressures in the Reference and H-junction designs, respectively, defined as $\Delta P_R^* = \Delta P_R / R_L Q_L$ and $\Delta P_H^* = \Delta P_H / R_L Q_L$.

Eqn. (7) highlights that an imbalanced flow condition ($Q_R \neq 1$), or a mismatch in the hydraulic resistances of the discharge channels ($R_D \neq 1$), can lead to parasitic pressure in both the Reference and H-junction designs. In contrast, the Y-junction design prevents the buildup of parasitic pressure downstream of the mixing microchannel by allowing both discharge streams to merge and flow together toward the outlet through a shared discharge channel.

Note that the hydraulic resistance ($R_h$) of a microfluidic channel is significantly influenced by both its geometry and the dynamic viscosity ($\eta$) of the fluid it transports[51]. For channels where the height ($h$) is less than the width ($w$)—a common characteristic in narrow mixing microfluidic channels—this resistance can be approximated as

$$R_h \approx \eta \frac{12L}{1 - 0.63\left(\frac{h}{w}\right)} \cdot \frac{1}{h^3 w} \quad \text{For } h < w, \quad (9)$$

where, $w$ refers to the width of the microfluidic channel. Given that the height of the mixing microchannels ($h$) is significantly smaller than that of the bypass and discharge channels ($H$), the hydraulic resistance of the mixing channels is substantially greater than that of the bypass

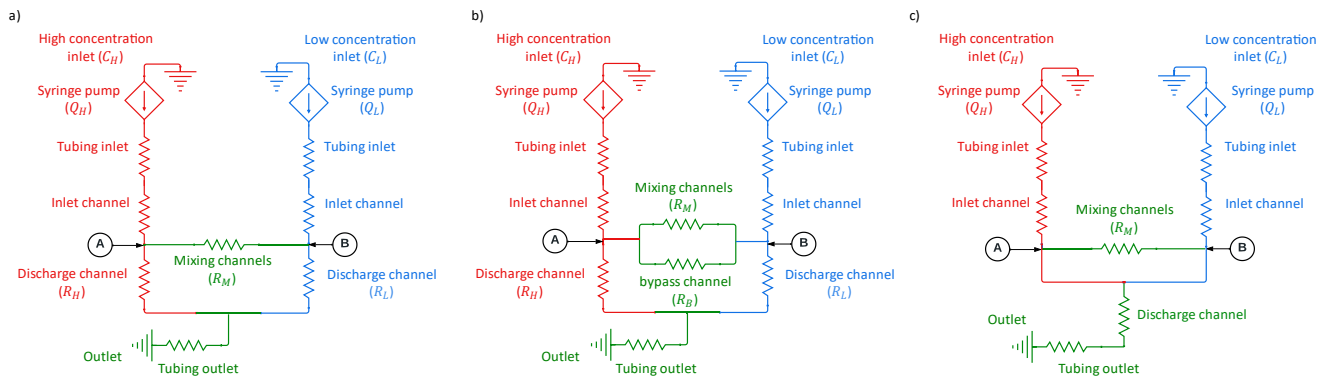

**Fig. 3**. Equivalent hydraulic circuit diagrams for **a**) the reference, **b**) H-junction, and **c**) Y-junction designs, illustrating parasitic pressure as the pressure difference between nodes A and B at the ends of the mixing microchannel. Syringe pumps introduce high- and low-concentration solutions ($C_H$ and $C_L$) at constant flow rates ($Q_H$ and $Q_L$), with microfluidic channels and tubing modeled as hydraulic resistors that impede flow. Hydraulic resistances for key components in the circuit analysis are annotated alongside their respective resistors.

Suppressing parasitic flow in membraneless diffusion-based microfluidic gradient generators and discharge channels, thus $R_M^* \gg R_D$ and $R_M^* \gg R_B^*$. In practice, resistances in our design differ by a factor of $10^6$.[52] The difference in resistances simplifies the ratio $R_M^*/(R_M^* + R_D + 1) \approx 1$. Consequently, the equation for parasitic pressure in the Reference design can be expressed as $\Delta P_R^* \approx (R_D Q_R - 1)$.

Similarly, for the H-junction design, the term $R_M^* R_B^*/(R_M^* + R_B^*)$ asymptotically approaches $R_B^*$, leading to an approximation of the parasitic pressure in the H-junction design as

$$\Delta P_H^* \approx \left(\frac{R_B^*}{R_B^* + R_D + 1}\right)(R_D Q_R - 1). \quad (10)$$

Fig. 4a-c illustrates the derived solutions for $\Delta P_R^*$ and $\Delta P_H^*$ across various operational conditions and device geometries, with $1 < Q_R < 10^2$, $1 < R_D < 10^2$, $10^{-2} < R_B^* < 1$. As expected, the Reference design experiences significantly higher parasitic pressure than the H-junction design under both imbalanced flow ($Q_R \neq 1$) and discharge mismatch conditions ($R_D \neq 1$). This invariably results in $\Delta P_H^* \ll \Delta P_R^*$.

In addition, $\Delta P_H^*$ is markedly lower for $R_D \neq 1$ than for similar conditions with $Q_R \neq 1$. This effect especially pronounced when $R_D \gg 1$, indicates that the H-junction design is more effective under discharge mismatch conditions—such as differences in discharge channel geometry or when fluids of differing viscosities enter the discharge channels—than under condition of imbalanced flow ($Q_R \neq 1$).

To quantitatively assess the effectiveness of the H-junction and Y-junction designs in reducing parasitic pressure, we define their performance relative to the Reference design as

$$\gamma_H = 1 - \frac{\Delta P_H^*}{\Delta P_R^*} \quad \text{and} \quad \gamma_Y = 1 - \frac{\Delta P_Y^*}{\Delta P_R^*}. \quad (11)$$

Within this framework, $\gamma_H$ and $\gamma_Y$ represent the design performance of the H-junction and Y-junction designs, respectively.

Given that $\Delta P_H^*$ is consistently lower than $\Delta P_R^*$, $\gamma_H$ ranges between 0 (as ineffective as Reference design) and 1 (full suppression of parasitic pressure). As evidenced Fig. 4d, $\gamma_H$ remains stable across varying $Q_R$ values (holding $R_B^*$, $R_D$, and $R_M^*$ constant) but increases with greater $R_D$ values (keeping $R_B^*$, $Q_R$, and $R_M^*$ constant). Furthermore, a lower relative hydraulic resistance of the bypass channel compared to the discharge channels, $R_B^* = R_B/R_L$, leads to an increase in $\gamma_H$. Notably, $\gamma_H$ achieves its peak for a bypass resistance $R_B^* \approx 0$, suggesting that in this case, the H-junction design approaches the design performance of the Y-junction design (i.e., $\Delta P_Y^* \approx 0$ and $\gamma_H \approx 1$). This result implies that $\Delta P_Y^* < \Delta P_H^* \ll \Delta P_R^*$, and therefore $\gamma_H < \gamma_Y = 1$ holds under both imbalanced flow and discharge mismatch conditions. Thus, the Y-junction design demonstrates a superior ability to reduce parasitic pressure compared to the H-junction design.

### Finite Element Simulation

Building on insights gained from hydraulic circuit analysis, a three-dimensional finite element calculation (COMSOL Multiphysics) was conducted to evaluate the effectiveness of the introduced designs in suppressing parasitic pressure flows. We modelled concentration gradient profiles and the development of parasitic pressure flows within the mixing microchannels of the H-junction and Y-junction designs, compared to the Reference design, which lacks mechanisms for reducing parasitic pressure flow. First, laminar aqueous flow was modelled by solving the Navier-Stokes equations in the microchannel geometries. Subsequently, diffusion of suspended particles in this flow was evaluated by solving the drift-diffusion equation. Diffusivity values within the range of $10^{-14} < D < 10^{-9}$ m$^2$/s were considered, corresponding to large macromolecules with diffusivities around $D = 10^{-13}$ m$^2$/s and small ions (e.g, Na$^+$) with $D = 10^{-9}$ m$^2$/s.[53,54]

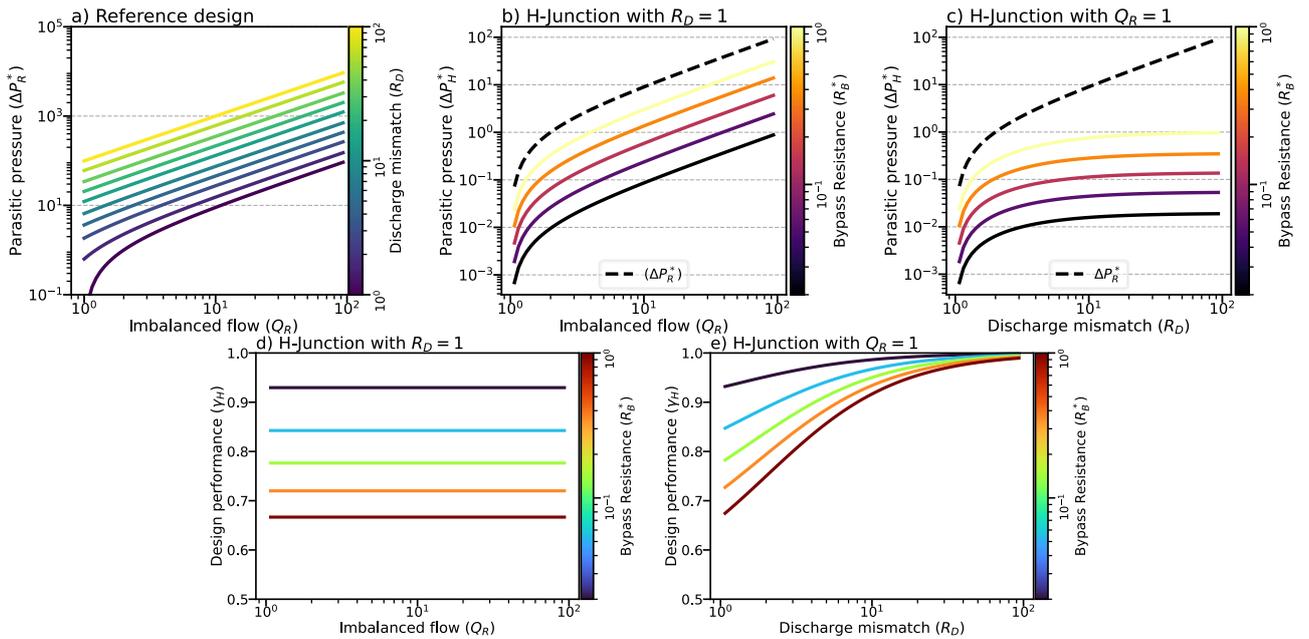

**Fig. 4**. Hydraulic circuit analysis results. **a)** Parasitic pressure generation in the Reference design ($\Delta P_R^*$) induced by imbalanced flow ($Q_R \neq 1$) and discharge mismatch ($R_D \neq 1$) conditions. **b, c)** Reduced parasitic pressure in the H-junction design ($\Delta P_H^*$) under imbalanced flow ($Q_R \neq 1$) in b) and discharge mismatch ($R_D \neq 1$) conditions in (c). **d, e)** Performance of the H-junction design ($\gamma_H$) in reducing parasitic pressure due to imbalanced flow ($Q_R \neq 1$) in (d) and discharge mismatch ($R_D \neq 1$) in e).

Suppressing parasitic flow in membraneless diffusion-based microfluidic gradient generators

To emulate imbalanced flow conditions, various inlet volumetric flow rates were examined, with $1 \leq Q_R \leq 100$. The simulations also covered a wide range of inflow concentrations to ensure that the normalized concentration gradients within the mixing microchannels, as defined by eqn. (4), remained consistently independent of the boundary conditions. Furthermore, we investigated various geometric configurations, specifically by adjusting the relative heights of the microchannels $h$ compared to the inlet/discharge side channels $H$, with $h/H \leq 0.025$, to take the impact of geometric factors into account.

Fig. 5 shows the results from this simulation, demonstrating the H-junction and Y-junction designs' effectiveness in minimizing parasitic pressure flows within the gradient region. Under identical geometric $(h/H)$ and operational $(Q_R)$ conditions, both the H-junction and Y-junction designs exhibited substantially lower levels of developed parasitic pressure and corresponding parasitic flow velocity compared to the Reference design. This observation aligns with the results from the hydraulic circuit analysis, suggesting that $\Delta P_Y^* < \Delta P_H^* \ll \Delta P_R^*$. Moreover, as depicted in Fig. 5c, the design performance parameter for the H-junction design ($\gamma_H$) is consistently lower than that of the Y-junction ($\gamma_Y$), indicating that $\gamma_H < \gamma_Y$. In addition, both $\gamma_H$ and $\gamma_Y$ are independent from the applied imbalanced flow conditions $Q_R$. These results validate the predictions of the circuit analysis, confirming that the Y-junction design is more effective than the H-junction in supressing parasitic pressure under imbalanced flow conditions.

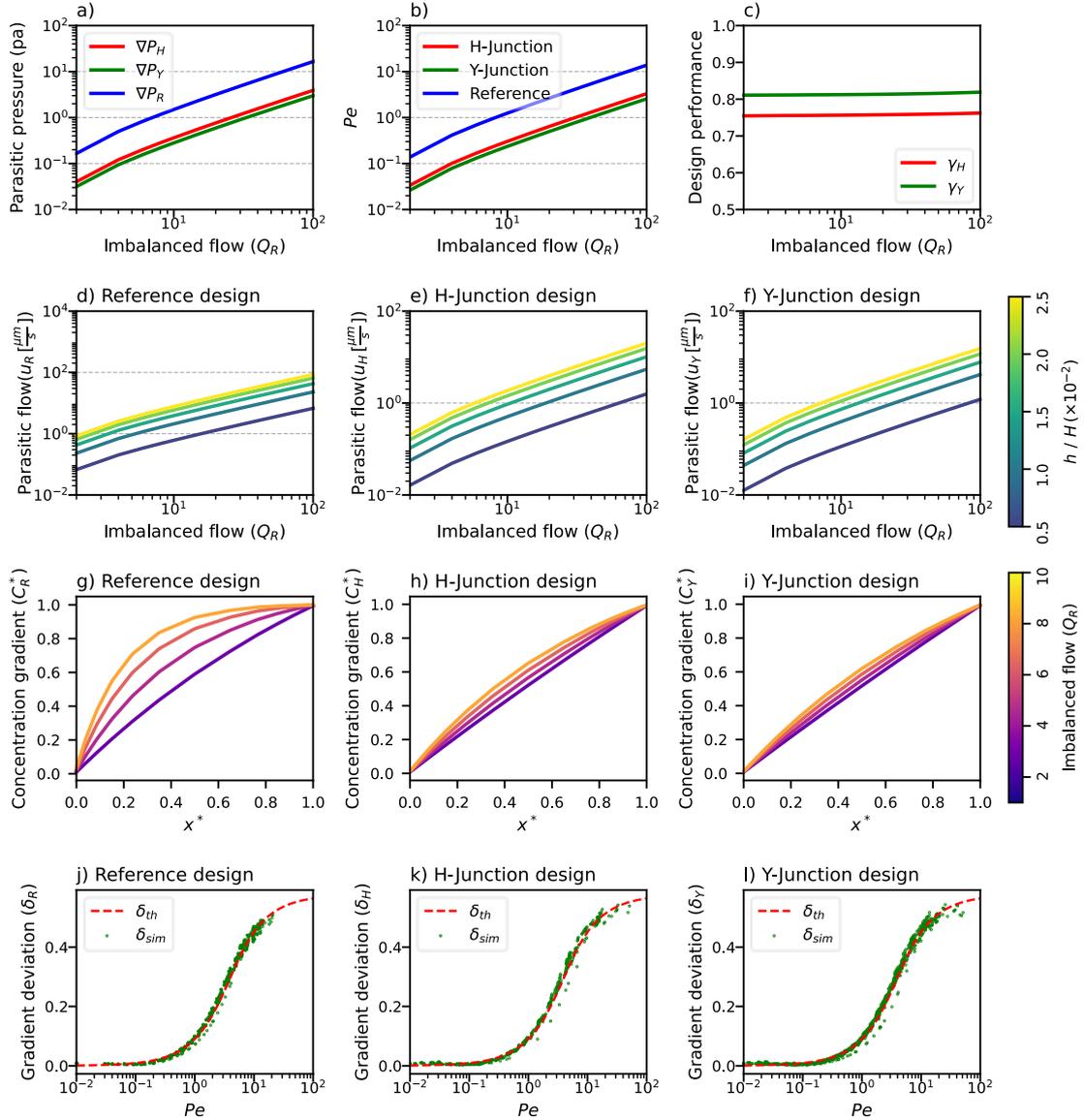

**Fig. 5**. 3D COMSOL simulation results on the mitigation of parasitic pressure flow by H-junction and Y-junction designs under imbalanced flow conditions ($Q_R \neq 1$). **a**) Parasitic pressure developed at the ends of mixing microchannels, and **b**) the resulting nonzero Péclet conditions with a height ratio $h/H = 0.01$ and diffusion coefficient $D = 1 \times 10^{-10}$ m$^2$/s. **c**) Design performance of the H-junction and Y-junction designs with $h/H = 0.01$. **d-f**) Mean parasitic flow velocity within the mixing microchannels, calculated across the microchannel's cross-sectional area. **g-i**) Deviation in concentration gradients from the linear gradient profile in the presence of nonzero Péclet conditions with a height ratio $h/H = 0.01$ and diffusion coefficient $D = 1 \times 10^{-10}$ m$^2$/s. **j-l**) $\delta$ parameter characterizing the deviation in concentration gradients derived from the fluid dynamic simulation ($\delta_{sim}$) and the theoretical model detailed in eqn. (6).

Suppressing parasitic flow in membraneless diffusion-based microfluidic gradient generators

However, noting that $\gamma_Y < 1$ contrasts with hydraulic circuit analysis, which predicted perfect pressure mitigation ($\gamma_Y = 1$) for the Y-junction design. This disagreement highlights the oversimplifications in the circuit analysis, particularly the assumption of flow convergence of discharge streams right after mixing channels. In experiment as well as in simulations, the inlet channels extend slightly beyond the mixing microchannel, and the hydraulic resistance of this extension was not considered in the analysis.

While parasitic pressure in DMGGs is influenced by the dimensions of the device downstream of the gradient region, its effect on parasitic flow in the mixing channel depends on the dimensions of the mixing channels. This highlights an alternative strategy for minimizing parasitic flow by employing notably small mixing channels (nanochannels), a method commonly employed in microfluidic research. This approach, while promising, is constrained by fabrication challenges and the spatial limitations of the gradient region, making it not consistently viable. In contrast, the innovative H-junction and Y-junction designs proposed in this study adeptly address this issue by directly diminishing parasitic pressure, and consequently, reducing parasitic flow within the gradient region.

Fig. 5d-f illustrates the generation of parasitic flow velocity because of parasitic pressure in the Reference, H-junction, and Y-junction designs, induced by imbalanced flow conditions ($Q_R > 1$). Notably, the parasitic flow velocity in both the H-junction ($U_H$) and Y-junction ($U_Y$) designs is substantially lower than in the Reference design ($U_R$), under the same operational conditions, signifying that $U_Y < U_H \ll U_R$. Additionally, flow velocities are decreased in channels with smaller height ratio $h/H$ in all three designs. Thus, reducing the dimensions of the mixing microchannels is a viable strategy for minimizing flow also in the new designs. However, the H- junction and Y-junction designs necessitate fewer dimensional modifications owing to the lower levels of parasitic pressure they generate, offering a more effective resolution to the challenge of parasitic flow in microfluidic systems.

Furthermore, deviations from the linear gradient profile, driven by parasitic flow, are distinctly evident as a function of the introduced imbalanced flow conditions as shown in Fig. 5g-i. H-junction and Y-junction designs exhibit significantly reduced deviations compared to the Reference design when subjected to the same imbalanced flow conditions. This indicates a reduced presence of parasitic flow within the gradient region for these new designs, enhancing the accuracy of determining local concentrations across the gradient domain based on preset concentrations $C_H$ and $C_L$. Additionally, the simulated deviations in gradient profiles, which are consistent with the analytical results depicted in Fig. 2a facilitate the precise calculation of the deviation parameter $\delta$ (specified in eqn. (5)). As shown in Fig. 5j-l the values of $\delta$ parameter derived from fluid dynamics simulations, denoted as $\delta_{sim}$, for the H-junction, Y-junction, and Reference designs align closely with the theoretical predictions $\delta_{th}$ calculated using eqn. (6).

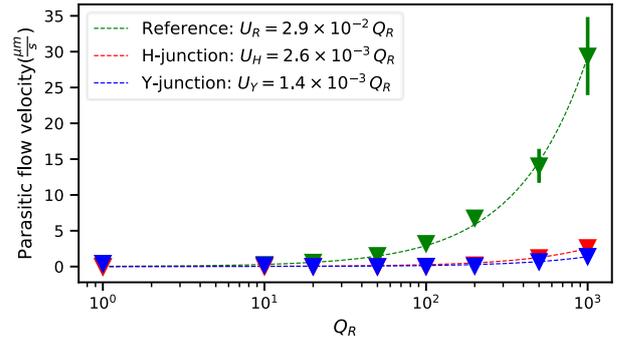

**Fig. 6.** Results from Particle Image Velocimetry analysis. Triangle markers illustrate parasitic flow velocity within the mixing microchannels across Reference ($U_R$), H-junction ($U_H$), and Y-junction ($U_Y$) designs subjected to imbalanced flow conditions ($Q_R \neq 1$). Linear fitting of the PIV data reveals a clear linear relationship (indicated by dashed fitted lines) between parasitic flow velocities and the applied $Q_R$.

**Particle Image Velocimetry**

Experimental studies were conducted to corroborate the simulation findings and to empirically evaluate the performance of the H-junction and Y-junction designs in mitigating parasitic pressure flow within the gradient region. For this purpose, Particle Image Velocimetry (PIV) was employed to measure flow velocities within the mixing microchannels under imbalanced flow conditions ($Q_R \neq 1$). The experimental setup included microfluidic devices configured with Reference, H-junction, and Y-junction designs, coupled with a high-resolution fluorescence microscope, and fluorescent particles of 200 nm diameter as tracers. An aqueous solution of 20 µg/mL tracer dissolved in Milli-Q water was introduced in both inlets, ensuring uniform distribution within the microchannels. Geometrical symmetry was maintained between both sides of the gradient generator (equal hydraulic resistance of channel on both side of the mixing channel), leaving the imbalanced flow ($Q_R \neq 1$) as the sole factor contributing to the development of parasitic pressure. Temporal variations in the tracer positions were captured using fluorescence microscopy, trajectories were determined through single particle tracking. These trajectories were analysed to calculate the average flow velocity, resulting from parasitic pressure developed at the ends of the microchannel, using mean-square-displacement (MSD) analysis.

PIV analysis results, indicating the parasitic flow velocities induced by imbalanced flow conditions ($Q_R$), are illustrated in Fig. 6. Under equivalent $Q_R$, the parasitic flow velocity in the Reference design ($U_R$) is significantly higher than those in the H-junction ($U_H$) and Y-junction ($U_Y$) designs, indicating $U_H \ll U_R$ and $U_Y \ll U_R$. This finding, further supported by Video S1 in Appendix, affirms the new designs' capability to substantially reduce parasitic flow. Moreover, the experimental results demonstrate that the Y-junction design is more effective in resisting parasitic flow generation than the H-junction design, as evidenced by $U_Y < U_H$ for identical $Q_R$ values, suggesting $\gamma_H < \gamma_Y$. These empirical observations align with the numerical results presented in Fig. 5. Additionally, a linear relationship emerged from analysing the experimental data, illustrating the relationship between the average parasitic flow velocities $U_R, U_H,$ and $U_Y$ and $Q_R$, further validating the simulation findings detailed in Fig. 5d-e.

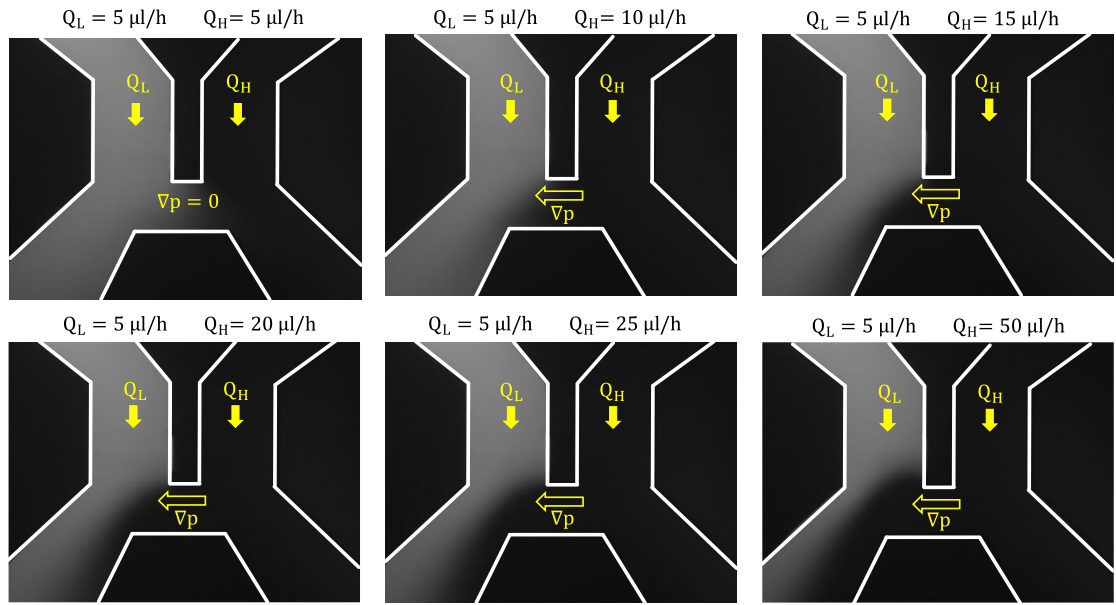

**Fig. 7**. Demonstrating the bypass channel's effectiveness in alleviating parasitic pressure ($\nabla p$) under imbalanced flow conditions ($Q_R = Q_H/Q_L > 1$). Milli-Q water flows through both sides of the gradient generator, with 10 mM fluorescein dye (FITC) introduced on the left side to mark the interface between the two laminar inflows. The ratio of volumetric flow rates for the left ($Q_L$) and right ($Q_H$) inlet channels are expressed as $Q_L : Q_H$ (μL/h), with the channel boundaries highlighted by white digital lines. (Mixing microchannel are present but not visible due to low fluorescence intensity in shallow channels.)

The performance of H-junction design in mitigating parasitic pressure was further validated by fluorescence microscopy. The results, presented in Fig. 7, show the process through which parasitic pressure is alleviated by the bypass channel located downstream of the mixing microchannel. A fluorescent dye was introduced to one of the inlet side channels to visually mark the flow interface, and volumetric inflow rates were adjusted to create imbalanced flow conditions ($Q_R \neq 1$). For $Q_R = 1$, no flow is detected through the bypass channel, suggesting no development of parasitic pressure ($\Delta p = 0$). However, this equilibrium state changes under conditions of imbalanced flow ($Q_R \neq 1$), where parasitic pressure ($\Delta p > 0$) is efficiently redirected through the bypass channel, minimizing its impact on the upstream mixing microchannels. Thus, this clear imaging directly shows efficient suppression of parasitic pressure and flow in the H-junction design.

## Conclusions

This study introduces new H-junction and Y-junction designs for membraneless DMGGs, aimed at suppressing parasitic flow within the gradient region. The effectiveness of these designs in minimizing parasitic pressure flow and providing a convection-free gradient was evaluated through hydraulic circuit analysis, 3D finite element simulations, and Particle Image Velocimetry. The findings confirm that these designs significantly minimize parasitic flow within the gradient region, achieving a concentration gradient profile that closely approximates the ideal pure-diffusion distribution. Additionally, these designs demonstrate enhanced performance under conditions of parasitic pressure caused by variations in side-channel geometry or fluid viscosity, positioning them as ideal for creating viscosity gradients[52,55] with the introduced concentration gradients.

Despite the effectiveness of both H-junction and Y-junction designs in countering parasitic pressure effects, the gradient region may still experience considerable parasitic flow under conditions of extreme flow imbalance, significant viscosity mismatch between side channels, or substantial differences in discharge channel dimensions. Such limitations may hinder the ability of these gradient generators to maintain a convection-free environment within the gradient region. To address this, we suggest exploiting counteractive effects to multiple sources of parasitic pressure flows, such as adjusting the inflow rate on the lower viscosity side to compensate for the higher viscosity's impact on the other side.

This research introduces innovative DMGGs that facilitate rapid gradient formation while maintaining a convection-free environment within the gradient region. By effectively minimizing parasitic pressure flows and the resultant shear stresses, these designs pave the way for new applications of membraneless DMGGs in future in-vitro studies. These applications include simulating biological microenvironments for single-cell research, developing precise and controllable transport systems for phoretic motion experiments, and facilitating concentration-dependent studies such as drug testing. Our gradient generators hold great potential for advancing research in various fields and enhancing the efficiency and accuracy of experiments that rely on concentration gradients

### Data Availability

The data sets and analysis code supporting the findings of this study are available in the Zenodo repository, accessible via the following DOI: https://doi.org/10.5281/zenodo.14024424.

Additional data can be made available upon request to the corresponding author.

Suppressing parasitic flow in membraneless diffusion-based microfluidic gradient generators

## Acknowledgements

We thank P.P.M.F.A. Mulder for technical support in the fabrication of devices, C. Richards for assistance in data analysis, and K.A. Sjollema for support with fluorescence microscopy.

Suppressing parasitic flow in membraneless diffusion-based microfluidic gradient generators

# Appendix

## Stability of boundary conditions

Maintaining stable boundary conditions (i.e., constant concentrations introduced to the ends of mixing microchannel) that govern concentration distribution within the mixing microchannel is crucial for generating accurate and reliable gradients using DMGGs. Parasitic flow and the concentration gradient inside the mixing microchannel result in variations in the preset concentrations at the microchannel ends, which must be compensated by introducing fresh solution into the side channels (inlet channels) to maintain constant concentration at the microchannel ends (boundary conditions). However, for extreme conditions including high parasitic flow, steep gradients, high diffusivity and low inflow rate, the boundary conditions may not be consistent with the preset concentrations.

As illustrated in Figure S1, the average random displacement at the microchannel ends towards the side channels over time $t$ equals

$$<\Delta x> = u_0 t - \sqrt{2Dt}, \text{ for } x = 0,$$
$$<\Delta x> = u_0 t + \sqrt{2Dt}, \text{ for } x = L. \quad (S1)$$

Here, $u_0$ represents the mean parasitic flow velocity, and $D$ denotes the diffusion coefficient. Consequently, the displacement within fluid replacement with refresh solutions ($t_w = \frac{w}{\overline{U}}$) will be $<\Delta x> = u_0(\frac{w}{\overline{U}}) \mp \sqrt{\frac{2Dw}{\overline{U}}}$, where $\overline{U}$ is the average parasitic flow velocity across the channel cross-section. Therefore, the minimum required $\overline{U}$ for the maximum allowable $<\Delta x>$ for both side channels can be expressed as

$$\overline{U}_{min} = \frac{w u_0^2}{(D + u_0 <\Delta x>_{max}) - \sqrt{D(D + 2u_0 <\Delta x>_{max})}}, \text{ for}$$
the right channel ($x = L$), and
$$\overline{U}_{min} = \frac{w u_0^2}{(D + u_0 <\Delta x>_{max}) + \sqrt{D(D + 2u_0 <\Delta x>_{max})}}, \text{ for} \quad (S2)$$
the left channel ($x = 0$).

Suppressing parasitic flow in membraneless diffusion-based microfluidic gradient generators

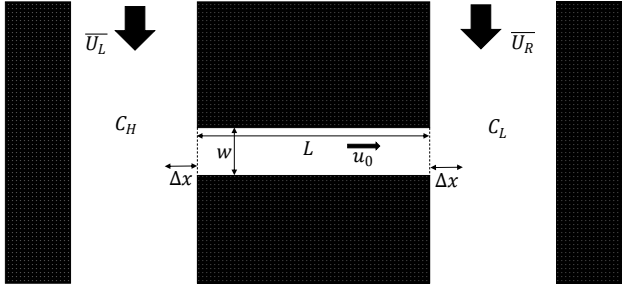

**Figure S1**. Schematic representation of a typical membraneless DMGGs: A concentration gradient is generated along a microchannel (with length $L$ and width $W$) connected to two relatively large side channels filled with two solutions having different concentrations, $C_H$ and $C_L$.

To minimize changes in the boundary conditions at the microchannel ends, the corresponding concentration change of a taken displacement of $<\Delta x>$ can be expressed as $\Delta c = <\Delta x>(\frac{\partial c}{\partial x})|_{x=0,L}$. Then, by including eqn (S1), the maximum allowable displacement $<\Delta x>_{max}$ can be determined for the permitted concentration change of $\Delta c_{permit}$ as

$$<\Delta x>_{max} = \frac{D\left(1 - e^{\frac{-u_0 L}{D}}\right)}{u_0(C_H - C_L)} \times \Delta c_{permit},$$
for $x = 0$,

$$<\Delta x>_{max} = \frac{D(1 - e^{\frac{-u_0 L}{D}})}{u_0(C_H - C_L)(e^{\frac{-u_0 L}{D}})} \times \Delta c_{permit}$$
for $x = L$.

(S 3)

Here, the permitted concentration change, $\Delta c_{permit}$ can be determined by defining a design parameter describing the reliability of the applied boundary condition respecting the uniform concentration inside the side channels ($C_B$) as $S = \frac{C_B - \Delta c_{permit}}{C_B}$. For instance, for a reliability of 99%, $\Delta c_{permit}$ at the left end of the microchannel will be $\Delta c_{permit} = C_H(1 - 0.99)$.

In summary, to ensure the stability of the applied boundary conditions, the inflow rates of a membraneless DMGGs must be set higher than the required inflow rates satisfying (S2).

**Particle Image Velocimetry**

A video is provided to complement the results of Particle Image Velocimetry (PIV) analysis illustrated in Fig. 6 in the main text. Link to video:
https://figshare.com/s/b81d49897570ce9a3e1e

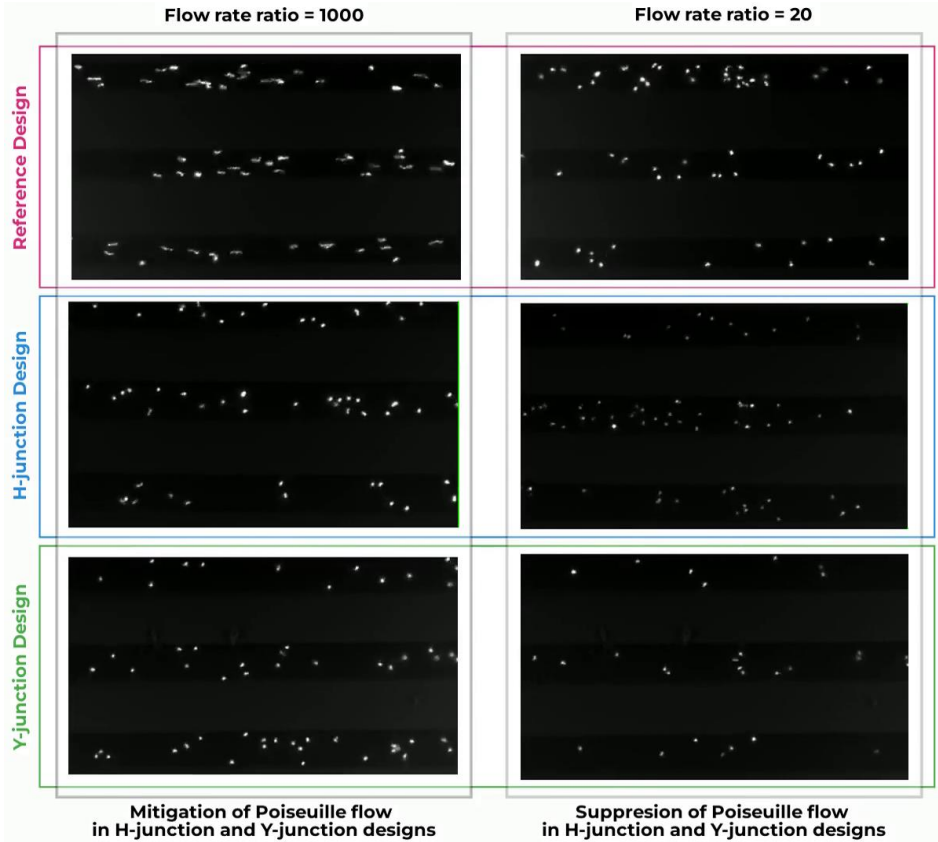

**Video S1**. Mitigation and suppression of parasitic flow within the microchannel using the H-junction and Y-junction designs. Polystyrene nanoparticles with a diameter of 200 nm suspended in Milli-Q water at a concentration of 50 µg/ml were introduced into both the left and right side-channels at constant flow rates of $Q_L$ and $Q_H$, respectively. In all experiments, $Q_L$ is consistently set to 5 $\mu L/h$. Depending on their column location (highlighted in light grey), the experiments have different $Q_H$ settings: those in the left column have a $Q_H$ of 5000 µL/h, resulting in a flow rate ratio $Q_R$ of 1000, whereas those in the right column have a $Q_H$ of 100 $\mu L/h$, yielding a $Q_R$ of 20. As visually demonstrated, the parasitic flow generated in the Reference design is substantially more pronounced than that in the H-junction and Y-junction designs when a flow rate ratio $Q_R$ of 1000 is applied (as indicated in the left column). Furthermore, at a lower flow rate ratio of $Q_R = 20$, parasitic flow is nearly undetectable in the H-junction and Y-junction designs, whereas it remains clearly visible in the Reference design. Link to video: https://figshare.com/s/b81d49897570ce9a3e1e.